\documentclass[12pt]{article} 
\usepackage{amssymb,latexsym,amsthm,amsopn,amstext,amscd,mathrsfs,amsmath}
\usepackage{amsfonts}

\usepackage[margin=3.3cm]{geometry}

\usepackage {graphicx}

\usepackage{authblk}

\theoremstyle{plain} 
\newtheorem*{dfn*}{Definition}

\begin{document}
 
\title{Gauge from holography and \\
holographic gravitational observables}

\author{	Jos\'e A. Zapata%
\footnote
{email: {\tt zapata@matmor.unam.mx}}
}
\affil{
Centro de Ciencias Matem\'aticas \\
Universidad Nacional Aut\'onoma de M\'exico \\
Morelia Mich. M\'exico
}

\maketitle

\begin{abstract}
In a spacetime divided into two regions $U_1$ and $U_2$ by a hypersurface $\Sigma$, 
a perturbation of the field in $U_1$ is coupled to perturbations in $U_2$ by means of the holographic imprint that it leaves on $\Sigma$. 
The linearized gluing field equation constrains perturbations on the two sides of a dividing hypersurface, 
and this linear operator may have a nontrivial null space. 
A nontrivial perturbation of the field leaving a holographic imprint on a dividing hypersurface which does not affect perturbations on the other side should be considered physically irrelevant. 
This consideration, together with a locality requirement, leads to the notion of gauge equivalence in Lagrangian field theory 
over confined spacetime domains.

Physical observables in a spacetime domain $U$ 
can be calculated integrating (possibly non local) 
gauge invariant conserved currents on hypersurfaces 
such that $\partial \Sigma \subset \partial U$. The set of observables of this type is sufficient to distinguish gauge inequivalent solutions. 
The integral of a conserved current on a hypersurface is sensitive only to its homology class $[\Sigma]$, and if 
$U$ is homeomorphic to a four ball the homology class is determined by its boundary $S = \partial \Sigma$. 
We will see that 
a result of Anderson and Torre implies 
that for a class of theories including vacuum General Relativity all 
local observables are 
holographic in the sense that they can be written as integrals of 
over the two dimensional surface $S$. 
However, non holographic observables are needed to distinguish between gauge inequivalent solutions. 

\end{abstract}

%
%
%

\section{Context}\label{ContextSect}
%

Motivated by the development of blackhole thermodynamics \cite{bekenstein1973black, bardeen1973four}, 
more than two decades ago pioneers of modern physics put forward a holographic principle that sparked immense interest in the community \cite{hooft1993dimensional, susskind1995world}. 
More recently, motivations from string theory lead to the 
discovery of the gauge/gravity correspondence \cite{maldacena} 
providing an avenue for defining new quantum gravity theories. 
Interestingly, 
quantum information has been found to play an important role in 
the gauge/gravity correspondence \cite{Ryu:2006bv, Liu:2013iza, 
Shenker:2013pqa, Donnelly:2017jcd}. 

Despite its motivation, this article 
does not contribute by elongating 
the promising road towards quantum gravity emerging from the gauge/gravity correspondence, and our work is classical in substance. 

A recurrent theme in this study is that a hypersurface locally splits spacetime into two regions and 
can be thought of as the communication channel between them. 

Spacetime localized properties of the field will be of our interest. Since measuring devices live in spacetime as well, measurement will be understood as the interaction between the system of interest (a certain field) and the measuring device (a field or a detector modeled otherwise). 
Think for example of a beam interacting with a screen for a short period of time; a spacetime description of the situation takes place in a bounded spacetime domain $U$ where part of its boundary is the world history of the screen. In this situation, our point of view will be that 
measurement did not take place inside a spacetime region $U$ where the field lives but at a boundary where it interacts with another system.
We will focus on the idea that measurement requires a division of a system into two subsystems, the system of interest and the reference system with respect to which we measure. 
A division of the field into two subsystems is achieved by considering a bounded spacetime region and considering the field inside $U$ as the system of interest and the field outside the region as the reference with respect to which we measure. 
The system of interest interacts with the reference system though the boundary of the region $\partial U$. We will exhibit observables for pure gravity defined on spacetime domains with boundary defined by considering the field at $\partial U$, 
together with its partial derivatives, 
as a reference system. 

From this point of view, there could be field configurations in $U$ which are different, but which cannot be distinguished by the reference system, which for us is the field outside of $U$ (or the field together with enough partial derivatives at $\partial U$). 

We will also play with the idea of having the ability of separating a system into subsystems in arbitrary ways as indicated by splitting a spacetime region into subregions like $U = U_1 \#_\Sigma U_2$ (which means that the union of the subregions is $U$ and the intersection is the hypersurface $\Sigma$). 
We will study the gluing conditions that the field and its perturbations satisfy at hypersurfaces $\Sigma$ serving as communication channel between subsystems. 

From the study of how perturbations propagate through communicating hypersurfaces a condition for gauge equivalence naturally arises. We complement it with requirements of locality and relativity of measurement to give rise to a notion of gauge vector fields which is suited to work on spacetime confined domains. 
This notion of gauge becomes a cornerstone for the version of Lagrangian field theory presented in this work and for the study of observables and holography given in the rest of the article. 

A close relative to our proposal is the 
discovery of holographic behavior in gauge theories confined to bounded domains; more precisely, 
that the presence of boundaries in gauge theories leads to 
``would be gauge degrees of freedom'' living at the boundary \cite{Freidel:2015gpa, donnelly2016local} 
showing that entanglement entropy, gauge and locality are 
interestingly intertwined \cite{donnelly2014entanglement, donnelly2016observables}. 

Our interest on communicating hypersurfaces $\Sigma$ and on measurement makes it natural to consider observables modeled by the integration of currents on hypersurfaces. We want to consider the data at $\Sigma$ as characterizing a solution (at least partially) in a neighborhood of $\Sigma$ and we want the function to depend on the solution being as independent of $\Sigma$ as possible. This is our reason for considering conserved currents. Additionally since our intention is to model measurement, we should consider gauge invariant conserved currents. In the paper we show that observables calculated integrating this type of currents are capable of distinguishing gauge inequivalent solutions.

In bounded spacetime domains $U$ with the topology of a ball 
observables calculated integrating conserved currents do not depend on more details of the integrating hypersurface than its boundary, 
$S=\partial \Sigma \subset \partial U$. The hypersurface can be deformed within $U$ keeping its boundary fixed and the evaluation of the observable  would not change. 
On the other hand, the field inside $U$ is completely determined by the field (together with enough of its partial derivatives) in $\partial U$. Generic observables in bounded spacetime domains 
can be thought of as complicated functionals of the field in $\partial U$. 

It is natural to wonder about observables calculated from conserved currents which depend only on the field at the co-dimension two surface 
$S=\partial \Sigma$. We 
study this type of holographic observables in detail.

Before we finish setting up the context of this study, we have to give a pair of remarks. The first one regards the causal structure. 
General relativity is one of the theories that we intend to cover in our study, and it sets spacetime geometry as a dynamical field interacting with matter fields. 
Since causal structure follows from spacetime geometry, 
at the initial stage of our setting we consider a spacetime $M$ without a 
prefixed causal structure which could let us talk about spacelike surfaces or Cauchy surfaces for example. 
Once we have a metric, or a class of metrics, we can refer to a causal structure. The previous remark gives us the opportunity to mention that we will consider observables which are not defined for all field configurations (or for all solutions). Important observables will need a specific physical context to be defined, and this context may only be consistent in a domain of definition consisting of a certain class of fields. A mathematical side to the same issue could be that the expressions defining a given observable may be well defined only in a certain domain. 

The second remark is about the term ``local''. In this introductory section we already mentioned 
``spacetime localized'' properties of the field; it refers to properties of the field inside a bounded spacetime region $U$. This term will not appear often and if it does its meaning should be clear. 
We also mentioned a principle of ``locality of measurement'' when we were talking about our notion of gauge equivalence. We will see later that this refers to being able to calculate a physical observable as a sum of ``local contributions'' where each of them is gauge invariant in the appropriate context. 
In the next section we will talk about ``local functionals'' of the field. This term will appear often and it refers to functionals depending on the field and finitely many of its partial derivatives. 
But the terminology needs to be more specific. 
A physical observable $f$ is called ``spacetime local'' if it is calculated integrating a density which depends on finitely many partial derivatives of the field. 
A physical observable $f_\Sigma$ calculated integrating a current 
is said to be ``hypersurface local'' if the current depends on finitely many partial derivatives of the field. 
Consider a field theory which admits a formulation in terms of initial data. Call $T_\Sigma$ the map sending initial data to solutions. We can use the pull back map $T_\Sigma^\ast$ to transform spacetime covariant functionals into functionals of initial data. 
Since this procedure involves integrating non-linear equations, its is generically a non-local procedure which transforms spacetime local functionals into functionals which fail to be hypersurface local. 
There are very few hypersurface local observables apart from Noether charges. 
A set of observables calculated integrating currents needs to include observables which are not hypersurface local to be rich enough to describe the system.


The organization of this paper is as follows. 
In Section \ref{LagrFT} we will present a version of Lagrangian field theory and pay special attention to the gluing conditions that the field and its perturbations satisfy at hypersurfaces $\Sigma$ separating the system of interest into two pieces. 
We chose a version of field theory based in two aspects which we considered essential for our study: spacetime covariance and locality. 
The version of Lagrangian field theory that we describe in the next section is a spacetime covariant and local description in the sense of Oeckl's General Boundary Formulation of field theory \cite{GBF}. 
The example of two dimensional abelian BF theory is developed in this section with the intention of providing a simple example which is useful for the rest of the article. General relativity is mentioned frequently in the paper, but instead of performing technical calculations for that example we give references for the results that we mention. 
In Section \ref{GaugeSect} we develop a notion of gauge equivalence appropriate for bounded spacetime regions. 
With this notion of gauge, the version of Lagrangian field theory presented here is completed. 
Our argument supporting the definition equivalence is a covariant and local version of Newton's principle of determinacy. 
In Section \ref{ObsFromCurrentsSect} we study the class of observables that can be calculated integrating gauge invariant conserved currents. 
In Section \ref{LocGrObsAreHolo} we describe how results of Anderson and Torre \cite{Torre:1993jm, Torre:1993fq} 
imply that hypersurface local gravitational observables (in the vacuum) are holographic. 
In Section \ref{ExamplesSect} we present families of gravitational observables which are holographic and large families of observables which are not holographic. 
Section \ref{ConclusionSect} contains a summary of our work and gives some  concluding remarks.

\section{Lagrangian classical field theory}\label{LagrFT}
In this section we give a brief review of the version of Lagrangian classical field theory that will used in this work. 
It will also be useful to fix the notation for the rest of the article. 
Local functionals and the variational principle 
are elegantly treated in the jet bundle using the tools of the variational bicomplex. 
Vector fields in the space of solutions play an important role in our work, in particular because observables and observable currents have associated Hamiltonian vector fields. In order to correctly model those vector fields we have to step out of the geometrical formalism of the jet bundle and work at the level of sections using analytical methods. 
Here we will give a minimal description of the part of this formalism that is essential for presenting our results. 
For an excellent short introduction the variational bicomplex see \cite{anderson1992introduction}, and for a combination of this framework with analytical methods see \cite{Forger+Romero, Khavkine-Peierls}. 
First we will have a general review of the formalism, and we will finish the section with the example of two dimensional abelian BF theory. 
We choose this example because it is very simple and it still let us illustrate how the formalism works for field theories with gauge freedom. In the paper we will often mention general relativity, but we will not do the calculations here and just cite the relevant references.

\subsection{General framework}
We will aim to have a {\em local description} of field theory closely related to the General Boundary formulation of field theory \cite{GBF}. 
In our setting there is a fixed spacetime manifold $M$ of dimension $n=4$, and we will study the field in a region $U \subset M$ which may have a boundary and corners. 

There is a bundle over spacetime $\pi : Y \to M$ 
with an $m-$dimensional fiber $F$, 
and the field studied in this work is considered to be a section of the restriction of the bundle to the region of interest, 
$\phi : U \to Y|_U$. 


Points in the $k-$jet bundle $\pi_{k,0}:J^kY\rightarrow Y$, $k=1,2,\dots$ are  
equivalence classes of local sections of $\pi$ that agree up to $k$-th order partial derivatives when evaluated at a given point $x\in M$. 
The evaluation of a section of the original bundle is written using a local chart as 
$\phi(x) = (x^1,\dots,x^i,\dots,x^n; u^1,\dots,u^a,\dots,u^m) \in Y|_U$. Accordingly, the $k$-jet gets 
the following coordinates 
\[
	\left(x;u^{(k)}\right):=
		(x^1,\dots,x^i,\dots,x^n;u^1,\dots,u^a,\dots,u^m;\dots,u^{a}_I,\dots) ,
			\in J^k Y |_U
\]
where $ i=1,\dots, n;\, a=1,\dots,m;$ and $I=(i_1,\dots,i_k)$ is a {\em multiindex} consisting on an unordered $k$-tuple of coordinate indices 
(because this type of indices indicate a $k$-th order partial derivative of a section). 
We write $|I|:=i_1+\dots+i_n=0,1,\dots,k$, $i_j\geq 0,i_j\in\mathbb{N}$ for the degree of the multiindex. 
(In the case $I=\emptyset $ we set $u_\emptyset^a=u^a$). 

Different jets are related by projection maps. 
The projection $\pi_{k+r,k}:J^{k+r}Y\rightarrow J^kY$ 
is defined by erasing the coordinates that do not fit in the jet of lower order. 
In this structure the infinite jet $J^\infty Y$ 
is the jet of highest order. That is, it is a space from which there is a projection to every jet of finite order; 
and one can think of any jet of finite order as a truncation of the infinite jet. 
This idea is formalized defining the infinite jet by an inverse limit. 
For notational convenience we  denote the infinite jet simply by $JY$. 

Smooth functions on $JY$ are those and only those that can be puled back from a jet of finite order. 
Vector fields on $JY$ are derivative operators on the ring of smooth functions. 

Given a section $\phi: U \subset M \rightarrow Y|_U$, its prolongation to the $k-$jet $j^k\phi:U \subset M \rightarrow J^k Y|_U$ is 
\[
	j^k\phi(x)=
		\left(x^1,\dots,x^i,\dots,x^n;
		\phi^1(x),\dots,\phi^m(x);\dots,
		\frac{\partial^{|I|} \phi^{a}}{\partial^{i_1} x^1 \dots \partial^{i_n} x^n}(x),\dots\right) . 
\]

In classical field theory the main objects are not points of $Y$ but sections of $Y$. Thus, curves of sections and 
flows of sections will be of primary importance. It turns out that 
given any section $\phi \in {\rm Hists}_U$ the tangent space to the space of sections $T_\phi{\rm Hists}_U$ can be 
properly modeled by objects (vector fields) 
coming from the jet, but generic vector fields in ${\rm Hists}_U$ cannot. 

A vector field in $Y$ generates a flow of points in $Y$. 
Vector fields sending sections into sections without moving points in the base manifold are vertical vector fields 
\[
V = 0 \frac{\partial}{\partial x^i} + V^a \frac{\partial}{\partial u^a} . 
\]
If the coefficients are allowed to depend on partial derivatives of the field up to finite order this ``vector field'' 
(called an {\em evolutionary vector field}) would 
generate a flow of sections that if seen in the jet $JY$ would be generated by the vector field in $JY$ 
\[
jV = \sum_{|I|=0}^\infty (D_I V^a) \frac{\partial}{\partial u^a_I} ,  
\]
where $D_i$ is the total derivative (to be defined below) and where $D_I$ implies a successive application of the total derivative as 
indicated by the multi index $I$. 
The important property of evolutionary vector fields is that they send sections in $JY$ which are prolongations of sections of $Y$ 
into other sections that are prolongations. 
The coefficients in the expansion of a vector field in $JY$ in terms of a basis 
are functions in the jet; that is, they may depend on arbitrarily many, but finite, partial derivatives of the field. 
When seen from the point of view natural to the infinite dimensional manifold ${\rm Hists}_U$ 
the vector fields obtained from vector fields in the jet belong to a very spacial class; they are called local vector fields. 
At the end of this section we will mention that at the level of solutions ${\rm Sols}_U \subset {\rm Hists}_U$ 
generic vector fields are not local vector fields, and including non-local vector fields is essential for a physically appropriate treatment of 
classical field theory \cite{Kasha, Khavkine-Peierls}.

A convenient basis for differential forms in $JY$ is generated by wedge products of the set of generators 
$\{ dx^i , \vartheta^a_I \}$, where 
\[
	\vartheta^a_I:=
		du^a_I-\sum_{j=1}^nu^a_{(I,j)}dx^j .  
\]
%
Factors of the type $dx^i$ are called  ``horizontal'' and factors of the type 
$\vartheta^a_I$ are called ``vertical''. 
The space of $p$-forms is a direct sum of spaces ${\mathsf{\Omega}}^{r,s}(J Y)$ which are products of exactly $r$ horizontal one forms and $s$ vertical one forms. 
Since the differential brings up the degree of forms by one, the direct sum structure makes the differential split as a sum of operators 
\[
\mathsf{d}= \mathsf{d_h} + \mathsf{d_v} , 
\]
where 
$\mathsf{d_h}: {\mathsf{\Omega}}^{r,s}(J Y) \to {\mathsf{\Omega}}^{r+1,s}(J Y)$ and 
$\mathsf{d_v}: {\mathsf{\Omega}}^{r,s}(J Y) \to {\mathsf{\Omega}}^{r,s+1}(J Y)$ are characterized by their action on functions 
\[
	\mathsf{d_h} f = 
		\left( \frac{\partial f}{\partial x_i}
		+
			u^a_{(J,i)}\frac{\partial f}{\partial u^a_J} \right) dx^i
			= (D_i f) dx^i , \quad \quad
			\mathsf{d_v}f
	=
		\frac{\partial f}{\partial u^a_I}\vartheta^a_I .
\]
The differentials of the generators are 
\[
\mathsf{d_h}dx^i = 0 , \quad \mathsf{d_v}dx^i = 0 , \quad 
\mathsf{d_h}\vartheta^a_I = dx^i \wedge \vartheta^a_{(I,i)} , \quad 
\mathsf{d_v}\vartheta^a_I = 0 . 
\]
The following identities are easy to verify 
\[
	\mathsf{d_h}^2=0 ,\qquad 
	\mathsf{d_v}\mathsf{d_h}= - \mathsf{d_h}\mathsf{d_v} ,\qquad 
	\mathsf{d_v}^2=0	.
\] 
\[
\iota_X \mathsf{d_h} F = - \mathsf{d_h} \iota_X  F  , \quad 
j\phi^\ast \mathsf{d_h} G = d \, j\phi^\ast G , 
\]
where $X$ is any evolutionary vector field, 
$\phi$ is an arbitrary section of $Y$, 
$F$ is a differential form of 
horizontal degree $k$ and vertical degree $r$, 
$G$ is a differential form of 
horizontal degree $k$ and vertical degree $0$ and 
$d$ stands for the de Rham differential in $M$. 
In the closely related context of the covariant phase space formulation of classical field theory \cite{lee1990local, Forger+Romero}
the counterpart of $\mathsf{d_v}$ is the field differential (or variational differential), and 
$\mathsf{d_h}$ leads to the spacetime differential. 

Let us consider Hamilton's principle of extremal action for a Lagrangian density of $k$th order, 
$L (j \phi(x)) = L(x^i, \phi^a(x), \partial_i \phi^\alpha(x), \partial_{I = (i_1, \ldots 1_k)}\phi^a(x))$. 
Given a history $\phi$, 
the variation of the action 
induced by the evolutionary vector field $V$ 
is written as 
\begin{equation}
dS_U [V_\phi]= \int_U j\phi^\ast {\mathscr{L}}_{jV} L = 
\int_U j\phi^\ast \iota_{jV} (E(L) + \mathsf{d_h} \Theta_L) , 
\end{equation}
where an essential result of calculus of variations 
says that integration by parts 
lets us write any differential form of type 
${\mathsf{\Omega}}^{n,1}(J Y)$, like 
$\mathsf{d_v} L$, uniquely 
as a source form%
\footnote{
A source form has horizontal degree $n$ and its vertical part is proportional to the generator 
$\vartheta^a$ which sees the variation of the field and not the partial derivatives of the variation. 
} 
 plus a boundary term,
$\mathsf{d_v} L = I(\mathsf{d_v} L) + \mathsf{d_h} \Theta_L$, and we have defined $E(L)= I(\mathsf{d_v} L)$. For a description of the integration by parts operator see \cite{anderson1992introduction}.

The field equation is written as $j\phi^\ast E(L) = 0$. Histories solving the field equation are called solutions 
$\phi \in {\rm Sols}_U \subset {\rm Hists}_U$. 
The space ${\rm Sols}_U$ is an infinite dimensional manifold; the tangent space of a solution $T_\phi {\rm Sols}_U$ is generated by variations of the field induced by evolutionary vector fields $V$ satisfying the 
linearization of the field 
equation around $\phi$. Our notation will be as follows: $V$ induces a tangent vector $v_\phi \in T_\phi {\rm Sols}_U$. 
In this language the conservation of 
the presymplectic current 
$\Omega_L = - \mathsf{d_v} \Theta_L$ 
follows from $0 = \mathsf{d_v}^2 L = \mathsf{d_v} (E(L) + \mathsf{d_h} \Theta_L)$, and 
is written as the vanishing of the differential form 
$\mathsf{d_h} \iota_{jW} \iota_{jV} \Omega_L$ 
when restricted to points of the jet $j \phi(x)$ 
which are the image of a solution and 
when both inserted variations satisfy the linearization of the field 
equation around $\phi$. 
If we integrate $\mathsf{d_h} \iota_{jW} \iota_{jV} \Omega_L$ 
on a hypersurface $\Sigma$ we obtain the functional 
$\int_\Sigma j^\ast\phi \iota_{jW} \iota_{jV} \Omega_L$ 
measuring the presymplectic product of the variations induced by 
$V$ and $W$ which due to the conservation law is insensitive to deformations $\Sigma'= \Sigma + \partial B$. 

Vector fields in ${\rm Sols}_U$, assignments of tangent vectors to points, 
are not properly modeled by objets of the jet. 
It turns out that 
for non-linear PDEs generic vector fields in ${\rm Sols}_U$ are not-local. 
In particular, we will see later on in the paper that 
for vacuum general relativity Anderson and Torre \cite{Torre:1993jm, anderson1996classification} 
showed that the set of local vector fields do not generate all the non-trivial flows in Sols that are relevant to study field theory. Generic flows in ${\rm Sols}_U$ are generated by non-local vector fields in ${\rm Sols}_U$.

For example, if we are interested in using two vector fields $v, w$ in ${\rm Sols}_U$ to insert them in the presymplectic form to obtain 
the functional that calculates 
their presymplectic product, 
the calculations leads to the integral of a conserved current that is non-local. We will change our notation to accommodate for non-local objets; we will write 
\[
\omega_{L \, \Sigma}(v, w) [\phi]= 
\int_\Sigma \tilde{\Omega}_L(v, w)[\phi] , 
\]
where the presymplectic current 
evaluated on tangent vectors to a given solution 
is the same that we defined above, but if we insert tangent vectors which do not vary locally (when the solution is modified) the result is not a differential form in the jet. 

When dealing with non-local objects (like non-local vector fields or non-local currents) we will work at the level of sections and modify the notation as done with the functional written above. 
Thus, symbols with a tilde denote $n-1$ forms in $M$ which depend on the field (possibly in a non-local way) and on (possibly non-local) vector fields. 
An important remark is that
we do not need to define the objects with tildes separately because 
if a given solution is chosen with the aim of evaluating the functional, we can work with local objects using the formulas written above using the language of the jet.

According to the terminology that we declared in the introductory section, the functional $\omega_{L \, \Sigma}(v, w)$ is not hypersurface-local. 

\subsection{Example: two dimensional abelian BF theory} \label{Ex}

This example will take place considering that spacetime is a cylinder, 
$M = {\mathbb R} \times S^1$. We can consider that the region of interest is 
$U= [t_i, t_f] \times [\theta_i, \theta_f]$. 

The fields that we will consider are a $1$-form in $M$, which we call $A$, and a $0$-form, which will be denoted by $B$. 
The field bundle is $Y = T^\ast M \oplus M \times {\mathbb R}$. 
We could continue calling the total field 
$\phi$, but we choose to change to a more descriptive symbol. A history in $U$ will be denoted by $\AE:U \to Y|_U$. 
We choose a chart in $U$ with coordinates $(t, \theta)$, and in $Y$ we denote points using the induced local trivialization. Then the evaluation of a section may be written as 
$\AE(t, \theta) = (t, \theta ; A_t(t, \theta), A_\theta(t, \theta), 
B(t, \theta))$. We may also use a more compact notation 
$\AE(x) = (x^i ; A_i(x), B(x))$

The evaluation of the prolongation of the section in the jet is written as 
$j\AE(x) = (x^i ; A_i(x), B(x); v_{ij} = \partial_j A_i(x), v^B_j = \partial_j B(x); \ldots)$. 
The curvature of $A$ is denoted by $F$ and in the local trivialization it is written as $F_{ij} dx^i \wedge dx^j= \frac{1}{2} (v_{ji} - v_{ij})dx^i \wedge dx^j$. When pulled back to $M$, the curvature yields a two form which is determined by a function specifying the 
proportionality factor with respect to the area form $dt \wedge d\theta$ 
given by the chart. 

The basis for tangent vectors in the jet induced by the chart can be written as  
$\{ \partial_i = \frac{\partial}{\partial x^i} , 
\partial^j = \frac{\partial}{\partial A_j} , 
\partial_B = \frac{\partial}{\partial B} ; 
\partial^{ji} = \frac{\partial}{\partial v_{ji}} , 
\partial_B^{j} = \frac{\partial}{\partial v^B_j} ; 
\ldots \}$. 
The basis of one forms adapted to the vertical horizontal decomposition is 
$\{ d x^i , 
\vartheta_j= d A_j - v_{ji} dx^i , 
\vartheta^B= d B - v^B_i dx^i ; 
\vartheta_{ji}= d v_{ji} - v_{jik} dx^k , 
\vartheta^B_i= d v^B_i - v^B_{ik} dx^k ; 
\ldots \}$. 
Notice that we have simplified the natural notation in the sense that we could have decided to write $\vartheta^A_j$ instead of the simplified symbol $\vartheta_j$, and the same for all the vertical generators related to partial derivatives of the $A$ field. Then we are using a notation in which the presence of a super index $B$ means that it is related to the $B$ field and the absence of a superscript means that it is related to the $A$ field.

We live as an exercise to the reader to calculate the horizontal and vertical differentials of the coordinate functions that we chose for $JY$. 
The most relevant in the calculations below are: 
$\mathsf{d_v} B = \vartheta^B , \, \mathsf{d_v} F_ij =  \frac{1}{2} (\vartheta_{ji} - \vartheta_{ij}), \, 
\mathsf{d_h} B = v^B_j dx^j , \, \mathsf{d_h} \vartheta_i = dx^j \wedge \vartheta_{ij}$. 

We consider the first order action 
\begin{equation}
S_U[\AE] = \int_U j\AE^\ast B F. 
\end{equation}
This is the two dimensional version of a class of field theories whose quantization was called ``quantum cohomology'' by Horowitz in the paper \cite{bf} 
where he introduces BF theories as non-abelian generalizations of this class of field theories. 

We start calculating $\mathsf{d_v} L$; we obtain 
\[
\mathsf{d_v} L = (\vartheta^B F_{ij} +  \frac{1}{2} B(\vartheta_{ji}- \vartheta_{ij})) dx^i \wedge dx^j . 
\]
After integration by parts
(to get rid of differential forms which see partial derivatives of the variation) we obtain 
\[
\mathsf{d_v} L = ( v^B_i \vartheta_j + F_{ij} \vartheta^B) \wedge dx^i \wedge dx^j + 
\mathsf{d_h} (-B \vartheta_j \wedge dx^j ) . 
\]
From this calculation we can read the field equations 
(which turn out to be linear) 
\begin{equation}
d B = 0 , \quad  d A = 0 , 
\end{equation}
and also the presymplectic current 
\begin{eqnarray}\label{OmegaBF}
\Theta_L &=& -B \vartheta_j \wedge dx^j , \\
\Omega_L &=&  \vartheta^B \wedge \vartheta_j dx^j . 
\end{eqnarray}

The space of solutions in this example is formed by pairs consisting of a constant function (the $B$ field) and a closed $1$-form (the $A$ field). 
Notice that since the field equations are linear, perturbations of the field obey the same equations. 

This example is completely different from general relativity in the sense that vector fields in the space of solutions can be modeled by local vector fields. The field equations and the linearized field equations are linear, and every class of solutions modulo gauge contains local representatives. Then if we work with only local fields and local vector fields the description continues being physically appropriate. 

In the next section we will be able to comment on the space of solutions modulo gauge. Then it will be clear why is that Horowitz called the quantization of this class of theories quantum cohomology.

\section{Gauge from holography}\label{GaugeSect}

Consider a solution of the field equations $\phi$ and a hypersurface $\Sigma$ separating spacetime into two connected components $U_1$ and $U_2$ that intersect along $\Sigma$. Clearly, if we restrict the solution to one of the components we get a solution in a restricted domain $\phi_i= \phi|_{U_i}$. 
Consider a different solution $\phi'_1$ over subdomain $U_1$ such that its restriction to $\Sigma$ coincides with $\phi_1$ up to its 
partial derivatives of order $k$ and recall that the field equation is of order $k+1$. 
It is fact that we could 
cut and paste to replace the portion of solution over $U_1$ with $\phi'_1$ and generate a different solution 
$\phi' = \phi'_1 \#_\Sigma \phi_2$. 
As far as the field over $U_2$ is concerned, 
the cut and paste operation is not physically relevant. 
In the case in which the field in $U_1$ ---the system of interest--- is being studied through measurements at $\Sigma = \partial U_1 = U_1 \cap U_2$,the field configurations $\phi_1$ and $\phi'_1$ are not distinguishable by any measuring device. Therefore a natural extension to Newton's principle of determinacy to this scenario requires declaring $\phi_1$ and $\phi'_1$ as physically equivalent.

Let us study the cut and paste operation just described. 
Consider two domains intersecting at a hypersurface $\Sigma$ in such a way that they form a composite domain $U=U_1 \cup U_2$. 
If the field $\phi = \phi_1 \#_\Sigma \phi_2$ 
is considered as smooth at each subdomain but only continuous over $\Sigma$, the variation of the action for the composite domain 
$dS_U [W_\phi] = (dS_{U_1} + dS_{U_2}) [W_\phi]$ includes a term 
\begin{equation}\label{GlEq}
\int_{\Sigma} (j\phi_1^\ast - j\phi_2^\ast) \iota_{jW} \Theta_L . 
\end{equation}
A physical field in $U$ must be an extremum of the action 
for all variations vanishing at 
zeroth order over 
$\partial U$. 
That is, the allowed variations do not affect the evaluation of the field at the boundary, but they can induce changes in the partial derivatives of the field at the boundary. 
Given a history $\phi \in {\rm Hists}_U$, 
let us call $\mathbb{V}_\phi^{\partial U} \subset T_\phi {\rm Hists}_U$ 
the space of variations considered by Hamilton's principle of extremum action. 
As described in the previous section, we will model $T_\phi {\rm Hists}_U$ by evolutionary vector fields in the jet. 
We will study the conditions for extremality of the action in two steps corresponding to a decomposition of $\mathbb{V}_\phi^{\partial U}$ as a sum of two of its subspaces 
$\mathbb{V}_\phi^{\partial U} = \mathbb{V}_\phi^{\partial U_1, \partial U_2} + 
\mathbb{W}_\phi^{\partial U, U_1, U_2}$, 
where the first summand is the space of variations vanishing at 
zeroth order over 
$\partial U_1$ and $\partial U_2$ and the second summand is the space of variations 
vanishing at 
zeroth order over $\partial U$ and 
satisfying the linearization of the field equation (possibly with source if $\phi$ is not a solution) in the interior of the domains $U_1$ and $U_2$. 
Consider an arbitrary $V \in \mathbb{V}_\phi^{\partial U}$; it generically does not belong to $\mathbb{V}_\phi^{\partial U_1, \partial U_2}$ because it may not vanish at zeroth order over $\Sigma \setminus \partial \Sigma$. However since the 
restriction to $\Sigma \setminus \partial \Sigma$ of the zeroth order component of 
elements of $\mathbb{W}_\phi^{\partial U, U_1, U_2}$ is completely unconstrained, there is a (possibly not unique) 
$W \in \mathbb{W}_\phi^{\partial U, U_1, U_2}$ such that 
$V - W \in \mathbb{V}_\phi^{\partial U_1, \partial U_2}$. 

Thus, the extremaility of the action is equivalent to demanding 
first extremality with respect to variations in $\mathbb{V}_\phi^{\partial U_1, \partial U_2}$ 
and 
later extremality with respect to variations in 
$\mathbb{W}_\phi^{\partial U, U_1, U_2}$. 
We know that the first condition is equivalent to imposing the field equation in the interior of the domains $U_1$ and  $U_2$. 
The second condition 
does not impose any further restriction to the field in the interior of the domains $U_1$ and $U_2$, 
but it requires that (\ref{GlEq}) vanishes for every 
$W \in \mathbb{W}_\phi^{\partial U, U_1, U_2}$.%
\footnote{
Integration by parts is unnecessary if $\Theta_L|_\Sigma$ is already a source form, as happens with the usual definition. 
} 
%
%
The resulting gluing field equation of order $k$ 
demands that the $V$-momentum crossing $\Sigma$ coming from $U_1$ 
needs to match the $V$-momentum crossing $\Sigma$ coming from $U_2$.

Because of our interest in the type of relative measurements described in Section \ref{ContextSect}, 
we are particularly interested in the propagation of variations of the field through hypersurfaces. 
Consider a one parameter family of fields $\phi_t$ (starting at $\phi_{t=0} = \phi$ and 
generated by the flow of the vector field 
$v = v_1 \#_{\Sigma} v_2$) 
satisfying the field equation in $U_1$ and $U_2$ and solving the gluing problem over $\Sigma$ for each value of the parameter.%
\footnote{
Now we are considering vector fields in ${\rm Sols}_U$, and as we mentioned earlier, restricting to local vector fields could be physically inappropriate. Following the notation of Section \ref{LagrFT}, 
we will indicate vector fields in the space of solutions by lower case letters and they will be inserted in multilinear linear operators denoted by letters with a tilde over them indicating that they are defined at the level of sections and they may be non-local. Recall that we do not need to define the objects with tildes separately because 
if a given solution is chosen with the aim of evaluating the functional, we can work with local objects using the formulas written in the language of the jet. 
} 
Since (\ref{GlEq}) 
vanishes for any $W \in \mathbb{W}_\phi^{\partial U, U_1, U_2}$ 
when evaluated at 
$\phi_t = \phi + t v + O(t^2)$; then up to first order in the $t$ parameter 
$(\Theta_L(j(\phi + t (v_1- v_2)))|_\Sigma$ 
vanishes 
when restricted to $\mathbb{W}_\phi^{\partial U, U_1, U_2}$. 
Thus, the linearized gluing field equation for 
the vector field 
$v = v_1 \#_{\Sigma} v_2$ is the requirement that 
\begin{equation}\label{LGEQ}
((
\mathscr{L}_{v_1- v_2} \tilde{\Theta}_L) (w)) [\phi] |_{\Sigma} =
- \tilde{\Omega}_L (v_1 -v_2 , w ) [\phi] |_{\Sigma} , 
\end{equation}
vanishes 
for any $w \in \mathbb{W}_\phi^{\partial U, U_1, U_2}$. 
A more convenient statement is that 
for any solution $\phi$ and 
any $w$ satisfying the linearized field equations (\ref{LGEQ}) 
vanishes up to 
an exact differential linear in $w$.%
\footnote{
Our derivation of equation (\ref{LGEQ}) uses 
$\mathscr{L}_{x} = \iota_{x} \delta + \delta \iota_{x}$, 
where $\delta$ stands for the field derivative, and the property that 
$(v_1 - v_2)_\phi|_\Sigma = 0$ 
implies that $(v_1- v_2)_\phi$ is in the kernel of $\tilde{\Theta}_L|_\Sigma$. 
}

%


It is interesting to characterize pairs of perturbations $V_1$ and $V'_1$ in $U_1$ that would be compatible with the same perturbation $V_2$ on $U_2$. 
There are two conditions: 
The first one is a continuity requirement along $\Sigma$, stating that the perturbations evaluated at the hypersurface agree. 
This condition, written in terms of their difference $x = v'_1 - v_1$, is the zeroth order condition $x|_\Sigma = 0$. 
The second condition arising from the linearized gluing equation is that 
$\tilde{\Omega}_L (x , w ) [\phi] |_{\Sigma}$
be an exact differential for any 
solution $\phi$ and for any vector field $w$ in ${\rm Sols}_U$. 
This condition carries information about the perturbations in the bulk of $U_1$ because, 
even when it is a condition imposed at $\Sigma$, 
it is sensitive to directional derivatives of the perturbation in directions transversal to the hypersurface. 
Moreover, of the two conditions the second one is the only one capable of transmitting information about $x$
in the bulk from $U_1$ to $U_2$ through $\Sigma$. 
We will call $v|_\Sigma$ the {\em holographic imprint} of a perturbation; the second condition compares the holographic imprint 
of two perturbations and it decides if they are distinguishable or not. 
Notice that, this condition is linear in $X$ and it may have 
a non trivial null space. In that case, there would be some perturbations which are different in the bulk of $U_1$, but such that their 
difference 
$x = v'_1 - v_1 \neq 0$ cannot be resolved by this condition; those aspects of bulk perturbation are filtered out by 
the gluing condition at $\Sigma$ and do not couple to $v_2$.

Now let us consider a spacetime domain $U$ without any physical division into subdomains, and study the propagation of perturbations 
through thought hypersurfaces arbitrarily placed inside $U$. 
A vector field $x$ satisfying the condition $x|_\Sigma = 0$ for every hyersurface $\Sigma \subset U$ is the zero perturbation $x=0$. 
On the other hand, there may be non zero vector fields $x$ such that 
\begin{equation} \label{X}
\tilde{\Omega}_L (x , w ) [\phi] 
\quad \mbox{ is an exact differential }
\end{equation}
for any 
solution $\phi$ and any vector field $w$ 
in ${\rm Sols}_U$. 
%
Such a vector field 
hits a hypersurface $\Sigma \subset U$ leaving a {\em null holographic imprint}, one that 
necessarily looses all information transversal to the hypersurface, for {\em any} hypersurface.%
\footnote{
The conservation law for the presymplectic current described in 
Section \ref{LagrFT} implies that 
$\tilde{\Omega}_L (x , w ) [\phi]|_\Sigma$ is an exact differential for a given hypersurface $\Sigma$ if and only if it is also an exact differential for any homologus hypersurface $\Sigma' = \Sigma + \partial B$. 
}  
Notice that vector fields of this type form a vector space. 
If these perturbations are regarded as physically irrelevant, and the space of physical perturbations is considered to be 
the quotient vector space in which perturbations satisfying condition (\ref{X}) are identified with zero, then 
we could say that physical perturbations are those capable of transmitting holographic information.  
This is a mild type of holographic principle which in the context of this work, where spacetime geometry has not been fixed a priori, 
may be thought of as the covariant ingredient needed for Newton's principle of determinism. 
Elevating this to a definition 
is described in the title of this article with the phrase {\em gauge from holography}.%
\footnote{
In Section \ref{LocGrObsAreHolo} we will 
talk about local gravitational observables exhibiting holographic behavior; 
there our use of the term holographic is in the stronger sense meaning that 
the observables can be evaluated through integration over codimension two surfaces. 
}


Relativity and locality of measurement motivate the second condition needed in our definition of gauge perturbations. 
The simplest way to present it is recalling that we were searching for conditions for two variations of the field 
$v_1$, $v'_1$ on a spacetime domain $U_1$ to be indistinguishable to any field perturbation $v_2$ over the spacetime 
domain $U_2$ intersecting with $U_1$ at $\Sigma$. Now we will consider the situation in which we have one domain of interest $U \subset M$ 
with two perturbations $v$, $v'$ and we look for conditions that make those two perturbations indistinguishable to 
any perturbation $w$ of the field in the complement of $U$. 
Those indistinguishable perturbations will be called gauge related in $U$. 
Because perturbations in $U$ are judged with respect to perturbations in its complement, 
it makes sense to demand that gauge perturbations do not disturb the reference system. 
Then since $\partial U$ separates $U$ from its complement, 
we should demand that a gauge vector field $x$ act trivially in the boundary 
\begin{equation} \label{TrivialAtBdary}
(x_\phi)|_{\partial U} = 0
\quad \mbox{ for any solution.}  
\end{equation}

Now we comment on locality of measurement. 
Consider a situation in which spacetime $M$ has a Cauchy surface $\tilde{\Sigma}$ 
without boundary and where 
there is an observable of interest which is calculated as the integral of a gauge invariant current, 
e.g. 
$f_{\tilde{\Sigma}} [\phi] = \int_{\tilde{\Sigma}} \tilde{F}[\phi]$. 
We may calculate $f_{\tilde{\Sigma}} [\phi]$ as a sum of local contributions one of which corresponds to our 
spacetime domain of interest 
as $f_{{\Sigma}} [\phi] = \int_{{\Sigma}} \tilde{F}[\phi]$, where $\Sigma = \tilde{\Sigma} \cap U$. 
The locality condition amounts to requiring that for any gauge invariant current%
\footnote{
A current $\tilde{F}$ is declared to be invariant under vector field $x$ 
in the space of solutions if and only if 
${\mathscr{L}}_{x} \tilde{F}[\phi]$ is a pure divergence for any solution. 
} 
and for 
any hypersurface with $\partial \Sigma \subset \partial U$ integrals of the type $f_\Sigma [\phi]$ can be evaluated and are gauge invariant. 
The locality condition is satisfied if equation (\ref{TrivialAtBdary}) holds. 
If we do not impose that gauge transformations act trivially on the boundary we would end up in a situation in which 
$f_{\tilde{\Sigma}} [\phi]$ is a gauge invariant observable which may be calculated as a sum of local pieces, but in which the terms of the sum are not gauge invariant individually.


%
%

\begin{dfn*}[Gauge vector fields in $U$]
A vector field $x$ in ${\rm Sols}_U$ is declared to be a gauge vector field if and only if $x_\phi$ is in the null space of the linearized field equation for every hypersurface and 
every solution $\phi$, and it additionally vanishes at $\partial U$. 
More precisely, we demand that\\
\noindent
(i) $x$ satisfies equation (\ref{X}) and that \\
\noindent
(ii) $x$ satisfies equation (\ref{TrivialAtBdary}). 
\end{dfn*}

%

Isolated gravitational systems may be modelled over a spacetime domain of the type $U = \Sigma \times [0,1]$ 
with the boundary $\partial \Sigma \times [0,1]$ representing a world tube at spatial infinity (and possibly an inner boundary modeling a horizon). 
Our framework applies to any Lagrangian formulation of General Relativity, 
where it is known that all generators of diffeomorphisms $x$ induce perturbations such that 
$\tilde{\Omega}_L (x , w ) [\phi]$ 
is a pure divergence  \cite{ashtekar1990covariant}, which implies that $x$ satisfies Condition (i) of the definition of gauge vector fields. 
However, regarding variations that do not 
vanish at infinity as gauge 
is inappropriate since they modify the reference frame needed to define energy, momentum and angular momentum. 
Thus, 
preserving a reference frame at the boundary that may be used as a reference for measurements is 
another motivation for Condition (ii) of the definition of gauge.

The standard definition for gauge perturbations used in Lagrangian field theory (appropriate in manifolds with no boundary) is in terms of families of symmetries of the Lagrangian depending on a parameter that may be locally varying, and through Noether's second theorem this is related to ambiguities and over determination in the field equation (see \cite{lee1990local}). It is known that the standard definition implies our Condition (i), and there are works which have conjectured that Condition (i) may be equivalent to that definition 
(see for example \cite{reyes2004covariant}). 

It can be verified that this definition of gauge perturbations leads to a Lie subalgebra of the algebra of vector fields in ${\rm Sols}_U$ 
and to a notion of gauge equivalence among solutions; see \cite{OCs} for a proof.

Now it may be illustrative for the reader to go back to the two dimensional abelian BF theory example. We have already calculated $\Omega_L$ for that example in equation (\ref{OmegaBF}). It is not difficult to show that condition (\ref{X}) for the vector field in the space of solutions 
implies that it must generate transformations in the closed one form by exact forms, i.e. $A' = A +  df$ for some function of the base manifold $f$. 

Recall that in the example spacetime was a cylinder $M = {\mathbb R} \times S^1$. If the domain of interest is the whole spacetime we can ignore Condition (ii) for gauge vector fields. In this case the space of solutions modulo gauge is the two dimensional space parametrizing constant functions (the $B$ field) and cohomology classes of $1$-forms (the $A$ field). 
On the other hand, if there are boundaries in the region of interest $U$ the space of solutions modulo gauge is larger. 
The usual space of solutions modulo gauge would be recovered only after gluing all the regions of our subdivision of spacetime. 
In the next subsection we will see that Condition (ii) allows for the existence of nontrivial observables 
measuring $A$ degrees of freedom 
in bounded regions $U$ of spacetime.

\section{Observables from currents}
\label{ObsFromCurrentsSect}

It is natural to consider functions of the field calculated integrating currents over hypersurfaces. 
If the currents depend locally on the field and its partial derivatives those functions may be written as 
$f_\Sigma [\phi] = \int_\Sigma  j\phi^\ast {F}$; 
in the general case we use the notation 
\begin{equation}
\label{f}
f_\Sigma [\phi] = \int_\Sigma  \tilde{F}[\phi] . 
\end{equation}
This functional is gauge invariant if the hypersurface is such that 
$\partial \Sigma \subset \partial U$ and 
the current is gauge invariant in the sense that for every gauge vector field $x$ 
and every solution $\phi$ the Lie derivative of the current 
$\mathscr{L}_{x} \tilde{F}[\phi] $ yields a pure divergence. 
Moreover, 
a current is conserved if when evaluated on any solution 
the function does not depend on local deformations of the hypersurface 
of the type $\Sigma' = \Sigma + \partial B$. 
In this case the resulting observable is a function of the solution $\phi$ defined by the current 
(and the homology type of the surface).

A physical observable $f$ is spacetime local if it is calculated integrating a density which depends on finitely many partial derivatives of the field. 
A physical observable $f_\Sigma$ calculated integrating a current 
is said to be hypersurface local if the current $\tilde{F}|_\Sigma$ depends on finitely many partial derivatives of the field; 
for those currents we could write 
$\tilde{F}[\phi]|_\Sigma = j\phi^\ast F|_\Sigma$ for some differential form $F$ in the restriction of the jet to $\Sigma$. 
The locality of an observable current $\tilde{F}$ can also be determined by an almost locally Hamiltonian vector field $v$ associated to it; 
if $v$ is a depends on finitely many partial derivatives of the field then the current is local and the observable is hypersurface local. 
Consider a field theory which admits a formulation in terms of initial data. Call $T_\Sigma$ the map sending initial data to solutions. We can use the pull back map $T_\Sigma^\ast$ to transform spacetime covariant functionals into functionals of initial data. 
Since this procedure involves integrating non-linear equations, it is generically a non-local procedure which transforms spacetime local functionals into functionals which fail to be hypersurface local. 

There are very few hypersurface local observables, but we will see in this section that there are plenty of (possibly non-local) observables calculated integrating currents. One reason to suspect it is that 
$T_\Sigma^\ast$ transforms every spacetime observable into an observable depending on initial data and if the resulting observable is a current 
(calculated with a collection of maps $T_\Sigma^\ast$ labeled by the integrating hypersurface), 
it must be conserved.

Here we will consider this type of observables where 
the domain of definition of the current may be a proper open subset of the space of fields \cite{Khavkine-Obs}. 
Consequently, we consider physical observables which may be only defined in certain open domain of the space of solutions. 

Since in this work we will refer multiple times to (possibly non-local)  gauge invariant conserved currents defined in an open domain of the space of solutions, 
we will 
use the term {\em observable current} to refer to a current of this type. 

In order to make definitions a bit more concrete let us consider as an example two dimensional abelian BF theory as defined in Subsection \ref{Ex}. The field equation for the $A$ field is $dA=0$; then $A$ is a conserved current and the functional 
\[
hol_\Sigma [\AE] = \int_\Sigma A 
\]
is gauge invariant if $\partial \Sigma \subset \partial U$. 
That is, the functional $hol_\Sigma$ is defined in the space of solutions modulo gauge in $U$, and its value is independent of deformations of the integrating surface of the form $\Sigma' = \Sigma + \partial R$ with $R \subset U$. 

Notice that if Condition (ii) for gauge vector fields were not present we would not have a gauge invariant observables measuring $A$ degrees of freedom in bounded domains. 
In that situation, if we glued a collection of bounded domains $\{ U_I \}$ that covered spacetime $M= \mathbb R \times S^1$ we would discover that even when there were no observables measuring the $A$ field in each of the domains $U_I$, after gluing them all together an observable measuring the holonomy of $A$ around a non contractible loop in the cylinder would appear. 
In contrast, in our setting the mentioned physical observable in $M$ is the sum of local contributions which when considered in the bonded spacetime regions where they are defined are gauge invariant with respect to the appropriate notion of gauge (not with respect to the notion of gauge relevant when the region under study is the complete spacetime).

The resulting family of observables is a large family capable of distinguishing 
solutions which are not gauge related. 
Thus, the family of observables includes Noether charges for systems with 
simple Lagrangian symmetries 
whose generators are defined everywhere in the domain of the Lagrangian density, 
and it includes many more observables. 
Two aspects of our treatment are essential for 
proving separability of points in the space of solutions modulo gauge: 
The first one is properly modeling the space of variations of a given solution and the space of vector fields in ${\rm Sols}_U$ as discussed in Section \ref{LagrFT}. 
The second essential aspect is allowing observables which are not defined for all fields, 
but which are defined only 
in a certain open domain in ${\rm Sols}_U$. 
This will allow a tight correspondence between the differential of 
observable currents and almost locally Hamiltonian vector fields. 
A locally Hamiltonian vector field $v$ in $U$ 
is a 
vector field in ${\rm Sols}_U$ 
which preserves the multisymplectic form in the sense that 
for any solution $\phi$ and any two 
vector fields $w, z$ in ${\rm Sols}_U$ 
\begin{equation}\label{llhh}
(\mathscr{L}_{v} \tilde{\Omega}_L )(w , z ) [\phi] 
\quad \mbox{ is an exact differential vanishing at } \partial U . 
\end{equation}
Locally Hamiltonian vector fields in $U$ preserve the presymplectic forms 
$\omega_\Sigma (w, z) [\phi]= \int_\Sigma 
\tilde{\Omega}_L (w , z ) [\phi]$ 
defined using any hypersurface with $\partial \Sigma \subset \partial U$. 
An {\em almost locally Hamiltonian} vector field is a solution of the linearized field equation 
satisfying a version of equation (\ref{llhh}) in which the condition that the exact differential should vanish at $\partial U$ is not imposed. 
Thus, almost locally Hamiltonian vector fields may not preserve the presymplectic forms due to boundary terms. 
The importance of this type of vector fields for us is that given a solution $\phi$ 
we can model the tangent space of the space of solutions $T_\phi{\rm Sols}_U$ by 
vertical vector fields in the jet satisfying the linearized field equation which moreover 
are almost locally Hamiltonian; we drop the condition over $\partial U$ for the boundary term in equation (\ref{llhh}) 
because it could become an obstruction to generating $T_\phi{\rm Sols}_U$. 
There is an alternative to working with almost locally Hamiltonian vector fields, 
if the domain of interest is of the form $U = \Sigma \times [0, 1]$, 
it is endowed with a foliation by Cauchy surfaces $\Sigma_t$, 
and we are only interested in evaluating integrals on hypersurfaces belonging to the foliation. 
The alternative 
is to work with locally Hamiltonian vector fields using a 
version of equation (\ref{llhh}) in which 
the exact differential is required to vanish only over $\partial \Sigma \times [0, 1]$.%
\footnote{
This alternative does not make sense if the metric is not such that the leaves of the foliation are 
Cauchy surfaces. 
If the gravitational field is part of the system under study, using this alternative implies considering 
only some allowed fields. For a discussion see Appendix A. 
}

The linearized field equation is a non linear partial differential 
equation, which is linear only when 
not studied in the whole jet, but only on $j\phi(U)$ for a fixed solution $\phi$; thus
when the linearized field equation is considered in the jet, 
its solutions may exist only 
in a certain open domain. 
Since we are interested in providing conserved currents which are linked to given solutions of the 
linearized field equation, it is important for us to consider conserved currents which may be defined only on a certain open domain. 
Given any almost locally Hamiltonian vector field $v$ there is 
a (non empty) family of observable currents which have it as their associated 
almost Hamiltonian vector field. That is, for any solution of the field equation $\phi$ 
there is a family of observable currents such that 
for any vector field $w$ in ${\rm Sols}_U$ 
\begin{equation}
\label{HOCs}
(\mathscr{L}_{w} \tilde{F} + \tilde{\Omega}_L (v, w) ) [\phi]
\end{equation}
is an exact differential. 
Thus, the derivative of $f_\Sigma [\phi]$ in the direction of a perturbation $w$, 
which we will write as $\mathscr{L}_{w}f_\Sigma [\phi]$, is given by 
$-\omega_\Sigma (v, w)[\phi]$ except for a boundary term which would vanish if 
the integrating hypersurface has no boundary or if $w_\phi$ vanishes over 
$\partial \Sigma$. 

It turns out that, 
in field theories with local degrees of freedom, 
observable currents can distinguish gauge inequivalent solutions. 
A local version this statement is as follows: 
Consider any given solution $\phi$ and any given curve of solutions $\phi_t$ starting at $\phi_0 = \phi$ and 
determined by a vector field $w$ in ${\rm Sols}_U$ 
which is not a gauge vector field. 
We will see that 
\begin{equation}
\left.
\frac{d}{dt}\right|_{t=0} f_\Sigma[\phi_t] \neq 0 . 
\end{equation}
Now we sketch the main part of the proof leaving out a case whose treatment requires a more detailed calculation for Appendix B. 

There are two non exclusive possibilities 
for $w$ not being a generator of gauge transformations; the first one is that  
condition 
(\ref{X}) fails and the second one is that condition 
(\ref{TrivialAtBdary}) fails.
Consider now the case in which condition (\ref{X}) does not hold for $w$. 
Thus, there is a solution $\phi$, a 
vector field $v$ in ${\rm Sols}_U$ and a point $p$ 
in the interior of $U$ such that%
\footnote{
Now we will denote points in spacetime by the letter $p$ in order to avoid confusion with our notation for vector fields in the space of solutions. 
}
\begin{equation}
\label{conj}
\tilde{\Omega}_L(w, v)[\phi] (p) 
\quad \mbox{ is \underline{not} an exact differential. }
\end{equation}
At this stage, our argument needs a concrete statement for the properties assumed about 
field theories with local degrees of freedom. 
We assume that in a theory with local degrees of freedom 
all perturbations of the field which are not gauge generators 
have conjugated perturbations which are 
localized; more precisely, 
given (i) any point $p \in U$, (ii) any solution $\phi$ 
and (iii) a $w$ satisfying the condition stated in equation (\ref{conj}), 
we assume that 
there must be a hypersurface $\Sigma$ containing $p$ 
(in which we chose an auxiliary volume element $\mbox{vol}_\Sigma$) and a choice of $v$ 
vector field in ${\rm Sols}_U$ 
(which can be chosen to be 
almost locally Hamiltonian) 
such that equation (\ref{conj}) holds and where 
$(\tilde{\Omega}_L(w, v)[\phi])|_\Sigma = \lambda \mbox{vol}_\Sigma$ 
with the proportionality constant $\lambda$ satisfying 
$\lambda(p) > 0$ and 
$\lambda|_{\mathcal U} \geq 0$ 
where ${\mathcal U}$ is a neighborhood 
of $p$ contained in the interior of $\Sigma$ such that 
$v|_{\Sigma \setminus {\mathcal U}} = 0$. 
Then 
\begin{equation}
\left. \frac{d}{dt} \right|_{t=0} f_\Sigma[\phi_t] = 
\int_\Sigma 
\mathscr{L}_{w} \tilde{F}[\phi]  
= -\omega_\Sigma (w, v) > 0 , 
\end{equation}
which concludes the part of the proof that we give in the main body of the article. 

Previous works on similar approaches to classical field theory argue that 
there are not many physical observables arising from integrating conserved currents 
besides Noether charges (see for example \cite{Forger+Romero, Helein, kanatchikov1998canonical, Kijowski, Goldschmidt(1973)}). 
The definitions on which our work is based 
differ from those used in the cited literature in two aspects: First, we allow for non-local 
currents. Second, the currents that we consider may be defined only 
in a certain open domain in ${\rm Sols}_U$. 
The previous paragraph shows that the observable currents considered here 
are an interesting source of physical observables.

The case 
that was not covered in the proof given above 
is that relation (\ref{TrivialAtBdary}) fails, which means that there is a solution $\phi$ and a 
point $p \in \partial U$ such that $w_\phi(p) \neq 0$. Since the sub case in which (\ref{X}) fails was treated above, 
considering the case where (\ref{TrivialAtBdary}) fails and (\ref{X}) holds would conclude the proof; 
this is the case in which $W$ is not a gauge perturbation only because it fails to vanish over $\partial U$ 
--- a ``would be gauge'' perturbation. Since this part of the proof involves detailed expressions of the boundary terms 
we deal with it in Appendix B. 

Before closing this section 
we comment on the type of measurements of the bubble chamber, which are not properly modeled integrating currents. 
We consider that if the field of interest $\phi$ couples with the measuring field $\psi$ in such a way that 
gauge equivalent fields $\phi_A \sim \phi_B$ are resolved, it should be claimed that the measuring field significantly disturbes the system, and the actual field being measured is the one corresponding to the composite system. 
If the system is not considerably disturbed, the measurement induces a perturbation of the field satisfying the linearized field equation 
(which could be chosen to be almost locally Hamiltonian). In this scenario, 
the tools provided in this section show that 
an alternative equivalent description of the measurement is constructed integrating a conserved current.

\section{Hypersurface local gravitational observables are holographic} \label{LocGrObsAreHolo}

We saw that a perturbation of the field modeled by 
an almost locally Hamiltonian vector field $v$ 
is related  to a family of observable currents by equation (\ref{HOCs}). 
If we call an element of this family $\tilde{F}_0$ all the other elements of the family differ from it by 
``a constant'' $\tilde{F} = \tilde{F}_0 + K$, where an observable induced by the ``constant'' $k_\Sigma$ is ``field independent 
except for possible field dependence'' due to boundary terms; 
in particular, when $\partial \Sigma = \emptyset$ 
the observable $k_\Sigma$ is truly field independent. 
Another important aspect of the relation set up by equation (\ref{HOCs}) is that given an observable current $\tilde{F}$ 
there is a whole class of almost locally Hamiltonian vector fields related to it. 
The difference between any two such perturbations $x = v_1 - v_2$ satisfies equation (\ref{X}). 

The results described in the previous section show that the family of observable currents and perturbations 
satisfying equation (\ref{HOCs}) is capable of distinguishing gauge inequivalent solutions. 
This suggests that all observable currents may satisfy that equation for a given 
almost locally Hamiltonian 
perturbation, and 
the intuition turns out to be a fact. For a more thorough explanation see \cite{OCs}.

Consider 
observables calculated integrating conserved currents of the type 
$f_\Sigma$ as defined in equation (\ref{f}) 
in a bounded spacetime domain with the topology of a four ball. 
In such a domain the homology type%
\footnote{
Hypersurfaces $\Sigma$ and $\Sigma'$ are considered homologous if 
$\Sigma' = \Sigma + \partial B$ where $B$ is dimension $n$ manifold with boundary contained in 
the closed spacetime domain $U$. 
} 
of a hypersurface with $\partial \Sigma \subset \partial U$ 
is determined by the spacetime codimension two surface $S=\partial \Sigma$. 
Due to the conservation law, 
evaluation of this type of observables depends only on the homotopy type of $\Sigma$, 
and in this situation $\Sigma$'s homotopy type is determined by $S$. 
Additionally, after our notion of gauge excluded perturbations which do not 
propagate trough hypersurfaces, 
the field modulo gauge in the interior is determined by 
$j\phi|_{\partial U}$. Thus, 
the evaluation of an observable of this type 
is determined by $S$ and by $j\phi|_{\partial U}$. 
This result suggests studying an interesting subclass of these observables such that their evaluation depends on 
$j\phi|_{S}$ even if the evaluation cannot be achieved integrating 
a differential form on $S$. 
We will see below that in the case of General Relativity 
hypersurface local observables of the type $f_\Sigma$ 
can be evaluated as integrals of differential forms over $S=\partial \Sigma$.

Let us recall the terminology declared in the introductory section now that we have developed concepts and notation that let us be more explicit. 
A physical observable $f$ is spacetime local if it is calculated integrating a density which depends on finitely many partial derivatives of the field. 
A physical observable $f_\Sigma$ calculated integrating a current 
is said to be hypersurface local if the current $\tilde{F}|_\Sigma$ depends on finitely many partial derivatives of the field; 
for those currents we could write 
$\tilde{F}[\phi]|_\Sigma = j\phi^\ast F|_\Sigma$ for some differential form $F$ in the restriction of the jet to $\Sigma$. 
The locality of an observable current $\tilde{F}$ can also be determined by an almost locally Hamiltonian vector field $v$ associated to it; 
if $v$ is a depends on finitely many partial derivatives of the field then the current is local and the observable is hypersurface local. 
We also mentioned in the introduction 
that for a field theory 
admitting a formulation in terms of initial data, 
the integration of the field equations induces a map 
which generically sends 
local spacetime covariant functionals into functionals of initial data 
which fail to be hypersurface local.

Consider General Relativity on a spacetime domain $U$. 
The field describing the system is the spacetime metric $g_{ab}$ and perturbations are written as 
$g_{ab} \mapsto g_{ab} + h_{ab}$. A perturbation corresponding to a (possibly field dependent) 
diffeomorphism generator $X[g]$ has the form 
$h^X_{ab} [g] = \nabla_{(a} X_{b)} [g]$. 
It is a well known fact that for General Relativity perturbations corresponding to diffeomorphisms satisfy the 
linearized field equation and satisfy equation (\ref{X}); that is, their insertion in the presymplectic current 
yields a pure divergence (see for example \cite{ashtekar1990covariant}). 
According to our definition, the gauge generators of General Relativity in the given domain 
turn out to be vector fields generating diffeomorphisms whose restriction to $\partial U$ is the identity.

Then the observables related to a gauge vector field $x$ by equation (\ref{HOCs}) yield no information because they 
are field independent. However, if $x$ satisfies equation (\ref{X}) but does not vanish over 
$\partial U$ its associated observables $f_\Sigma^x[g]$ do carry non trivial information. 
Those observables 
may have a field dependence due to boundary terms 
\begin{equation}
f_\Sigma [g] = \int_\Sigma \tilde{F}^x[g] = 
\mbox{ const. } + 
\int_{\partial \Sigma} \tilde{\sigma}^x[g]  , 
\end{equation}
where $\tilde{\sigma}^x$ is a field dependent $n-2$ differential form. 
Even when these observables 
are initially written in terms of bulk fields, 
they can be calculated as integrals on the boundary of the integrating hypersurface; 
one may say that those observables 
are holographic.

Anderson and Torre \cite{Torre:1993jm, anderson1996classification}  
proved that the only solutions of the linearized field equation for vacuum General Relativity 
in the absence of matter 
$h_{ab}(jg(x)) = h_{ab}(x, g, \partial_x g, \ldots )$ depending on an arbitrary, but finite, 
number of partial derivatives of the metric are 
\begin{equation}
h_{ab} = h^X_{ab} + c \, g_{ab} . 
\end{equation}
Their result shows that, apart from rescaling generators, 
all {\em local} 
vector fields in ${\rm Sols}_U$ 
are gauge or ``would be gauge'' vector fields. 
Moreover, since rescaling generators are not almost locally Hamiltonian vector fields, 
those vector fields are not associated to gravitational observables. 
A corollary of Torre \cite{Torre:1993fq} 
can be read as saying that 
{\em all hypersurface local gravitational observables (in the vacuum) are holographic}. 

In the next section we will exhibit a family of holographic gravitational observables and a large family of non holographic gravitational observables. 

In contrast to the case of general relativity, as discussed in Subsection \ref{Ex} 
the example of two dimensional abelian BF theory is so simple that local fields and local vector vector fields are enough for a physically reasonable model of the system. The observable currents that we exhibited in Section \ref{ObsFromCurrentsSect} 
for that field theory are local and the induced observables $hol_\Sigma$ are not holographic.%
\footnote{
However if the domain of interest is not the cylinder spacetime but a domain $U$ with the topology of a ball, 
the closed form $A$ becomes exact when restricted to $U$, and 
the observables $hol_\Sigma$ are indistinguishable from observables which are holographic. 
}

Khavkine has proposed a generalized notion of spacetime local observables and exhibited a large family of gravitational observables which are spacetime local according to his definition \cite{Khavkine-Obs}. 
As mentioned previously, if those observables are transformed into functionals of initial data and those functionals turn out to be integrals of currents, they would generically be non-local observable currents.

\section{Examples of gravitational observables, holographic and not holographic} \label{ExamplesSect}

For the sake of concreteness, in this section we will consider a spacetime domain $U$ 
with the topology of a four ball.

There are many gravitational observables in our setting. 
According to the conventions stated in the previous paragraph, 
the components of the induced metric on the boundary (the pull back of the spacetime metric to the boundary) 
$q_{ij}^{\partial U}(p)$ evaluated any point $p \in \partial U$ 
are considered observables. 

This fact just talks about our choice of reference system. Part of our motivation to impose equation 
(\ref{TrivialAtBdary}) to gauge vector fields was to provide a reference with respect to which we could measure 
the system of interest ---the bulk field. The observables mentioned in the previous paragraph are not of physical interest 
in themselves, but they will enable us to define interesting gravitational observables. 

In Section \ref{ObsFromCurrentsSect} we described how in our framework 
given an almost locally Hamiltonian vector field which is not a gauge perturbation there is 
a family of observable currents corresponding to the given vector field. 
We also argued that for field theories with local degrees of freedom 
there are plenty of almost locally Hamiltonian vector fields 
which are not gauge.  
Since for any observable associated to a locally Hamiltonian vector field 
which is not gauge the second term in (\ref{HOCs}) does not vanish, 
those observables are not holographic.

In the previous section we asked if 
hypersurface non local gravitational observables could be approximated by local ones implying that 
non local gravitational observables are also holographic. 
With the aide of equation (\ref{HOCs}) we see that this question is equivalent to asking if 
for any given solution $\phi$ the tangent space to the space of solutions at $\phi$ can be generated by local 
almost locally Hamiltonian vector fields, 
where the term ``locally Hamiltonian'' refers to preserving the pre-symplectic form and the term ``local'' refers to 
depending on partial derivatives of the field of at most order $k$ for some finite $k$. 
In this terminology Anderson and Torre's result cited above says that the tangent space to the space of solutions 
of vacuum General Relativity 
based on any solution is not generated by the 
local solutions of the linearized field equation. Local perturbations generate only gauge or would be gauge directions, 
proving that hypersurface non local gravitational observables cannot be approximated by hypersurface local ones. 

A large family of observables is given by the symplectic product of physical perturbations (see \cite{OCs}). 
There is a subfamily which is holographic and a subfamily which is not. 
Assume that we are given two almost locally Hamiltonian vector fields $v, w$. 
The observable referred to as the symplectic product is 
\begin{equation}
\omega_{L \, \Sigma}(v, w) [\phi]= 
\int_\Sigma \tilde{\Omega}_L(v, w)[\phi] .
\end{equation}
It is simple to verify that this observable is associated to the  Hamiltonian vector field 
$[v, w]$. 
If $[v, w]$ satisfies equation (\ref{X}) and 
does not vanish in the jet over $\partial U$, then $\omega_{L \, \Sigma}(v, w)$ 
is a non trivial holographic observable. 
These are not very extravagant conditions; 
in the case of General Relativity, they would be satisfied by two diffeomorphisms generators 
whose commutator does not vanish over $\partial U$. 
A point that needs to be mentioned is that 
in General Relativity 
the diffeomorphism generators may be field dependent and in this way 
observables of 
this type encode non trivial information about the field. 
On the other hand, if $[v, w]$ does not satisfy equation (\ref{X}) 
the second term of (\ref{HOCs}) does not vanish implying that $\omega_{L \, \Sigma}(v, w)$ 
is not a holographic observable.

Below we give another family of examples of gravitational observables that can be defined thanks to the existence of a reference at the boundary. 
If the reader would like to have further motivation for considering bounded spacetime domains, 
it could be illuminating to read Appendix A. 
There we use a 
covariant initial value formulation on a given hypersurface with boundary $\Sigma$, 
which may be thought of as a laboratory where 
enough data is retrieved at a given time as to determine a solution inside $U = D(\Sigma)$, 
to discuss key aspects of the formalism described earlier in this article.

When the location of individual points inside $U$ can be determined in terms of 
matter fields, 
a wealth of options for gravitational observables opens up. 
Different variants of the procedure to construct observables of the system of gravity coupled to matter 
used at the beginning of our argument have appeared in uncountably many references. 
If two points $p_1, p_2 \in U$ (which are sufficiently close according to the metric field) 
are located using relations involving matter fields 
an example of an observable is the length of a geodesic $\gamma_{p_1 p_2}$
\begin{equation}
\mbox{Length}(\gamma_{p_1 p_2})[g , \phi] ,
\end{equation}
where we have denoted the gravitational field by $g$ and the matter fields collectively by $\phi$. 
If the gauge choice is changed, the location of $p_1$ and $p_2$ in the coordinate chart will change 
and the coordinate expression of the metric will also change in such a way as to 
leave the value of the observable unchanged; this is what makes it a physical observable. 
Notice that the condition that the points $p_1$ and $p_2$ be sufficiently close forces us to 
consider observables which are only locally defined in the space of fields. 
This observable depends on the 
the gravitational field 
and on the 
matter fields relative to which the points are located. 

On the other hand, if the points $p_1$ and $p_2$ belong to $\partial U$ they can also be determined by 
the matter fields (or even by the gravitational field) outside of $U$. If this is done, the family of observables 
$\mbox{Length}(\gamma_{p_1 p_2})$ 
has the interpretation of measuring the gravitational field in $U$ with respect to 
a reference frame determined by the fields outside of $U$; 
then we can proceed assuming that $\mbox{Length}(\gamma_{p_1 p_2})$ measures the gravitational field inside $U$ and 
is defined without the need of matter fields inside $U$. 
Furthermore, considering the 
restriction of the field to $\partial U$ 
as unaffected by changes of gauge (as demanded by equation (\ref{TrivialAtBdary})) 
implies that the function 
$\mbox{Length}(\gamma_{p_1 p_2})$ depending only gravitational field inside $U$ is an observable. 
The interpretation of such an observable is that 
it measures the field inside $U$ with respect to a reference system located at $\partial U$ 
which could be thought of as determined by fields outside of $U$.

Similar arguments could be used to define physical observables measuring areas of minimal surfaces determined by 
curves fixed at $\partial U$, or measuring volumes enclosed by collections of 
such surfaces. 

Are observables in these families holographic? 
They are not. 
One way to see that this might be true is to observe that 
the requirement for the functional to be an observable is that $\gamma_{p_1 p_2}$ 
is a {\em spacetime geodesic} 
and when spacetime data has to be retrieved from initial data at a given hypersurface the 
non linear field equations together with any gauge condition used enable calculations need to be solved 
to have the location of the spacetime geodesic determined in terms of the initial data. 
Torre's result  \cite{Torre:1993fq} confirms that our suspicion that $\mbox{Length}(\gamma_{p_1 p_2})$ is not a holographic observable is in fact correct.

\section{Conclusions}\label{ConclusionSect}

We started reviewing the notion of what a gauge perturbation is in the context of Lagrangian field theories 
defined on confined spacetime domains. 
The initial assumptions included that we are working in a covariant setting in which spacetime geometry may be one 
of the dynamical fields, implying that there is no causal structure fixed a priori. 
The initial consideration to determine which perturbations are considered gauge was 
a covariant form of a determinism principle thoroughly explained in Section \ref{GaugeSect}. 
The outcome was equation (\ref{X}), and it emerged from a study of how perturbations propagate through 
hypersurfaces as dictated by the linearized gluing equation. 
Since the transmission of information through spacetime codimension one surfaces played a key role, we referred to the argument as 
gauge from holography, and it was admitted that the term referred to 
a mild type of holography as compared to the 
the holographic behavior of hypersurface local gravitational observables described on Section \ref{LocGrObsAreHolo} 
which is associated to spacetime codimension two surfaces. 

At first glance, 
it may seem that both aspects of holography mentioned in the previous paragraph are disjointed, and that the term holography 
refers to completely are unrelated phenomena in both instances. 
However, the work presented in this article shows the intimate relationship between them: 
First of all, from our definition of gauge vector fields and our considerations of observables which are only 
locally defined in the space of solutions, we could give a straight forward argument proving that observables calculated 
integrating observable currents $f_\Sigma$ (see equation \ref{ObsFromCurrentsSect}) 
are capable of distinguishing gauge inequivalent solutions. 
A key ingredient of the argument was that almost locally Hamiltonian vector fields (these are perturbations of the field 
which respect the pre-symplectic form up to boundary terms which may not vanish) 
defined locally in the jet 
generate the tangent space to the 
space of solutions based at any solution. 
Second, any almost locally Hamiltonian vector field $v$ induces a (family of locally defined) observables $f^v_\Sigma$ 
which are holographic if $V$ satisfies equation (\ref{X}) and they are non trivial if condition 
(\ref{TrivialAtBdary}) is not satisfied. Thus, field perturbations related to non trivial 
holographic observables are exactly those 
which could be referred to as ``would be gauge vector fields'' in the sense that they are not gauge only due to the 
presence of a boundary. The information encoded in those observables can be non trivial because the 
perturbations satisfying equation (\ref{X}) ---in the case of General Relativity the diffeomorphism generators--- may 
depend nontrivially on the field. 

In Section \ref{ExamplesSect} 
we exhibited a large family of examples of observables corresponding to the symplectic product of field perturbations, 
and we characterized the subfamily consisting of non trivial holographic observables.

Observables in nonlinear theories with gauge redundancies 
defined in spacetime domains foliated by Cauchy surfaces with no boundary, 
are expected to be nonlocal in the sense of depending on 
infinitely many derivatives of the field. 
A brief discussion of the reasons behind this expectation 
is given in Section \ref{ExamplesSect}. 
However, we saw that in the presence of boundaries 
non trivial observables with holographic behavior arise.

Can every gravitational observable be approximated by local observables, inheriting their holographic behavior? 
This issue is addressed in Section \ref{ExamplesSect}, where we show that this is not the case.

Is the family of holographic observables capable of separating points in the space of solutions modulo gauge? 
The arguments given above tell us that this question is equivalent to asking if for any given solution the 
tangent space to the space of solutions based on it is generated by gauge vector fields together with would be gauge vector fields, 
which is clearly not the case.

\section*{Apendix A}\label{AppendixA}

In this appendix we motivate working with bounded spacetime domains from the point of view 
of a covariant initial value formulation on a given hypersurface with boundary which may be thought of as 
a laboratory where data is retrieved at a given time. We comment on the correspondence between this formalism and the 
one used in the main body of this article emphasizing 
interpretational issues which arise when setting up the correspondence and their relation with two key 
aspects of our formalism: 
considering locally defined observables and 
the notion of gauge.

For the sake of concreteness, 
consider the following scenario for gravitational 
thought measurements on earth: Earth's southern hemisphere 
is covered by laboratories 
(covering a layer from a height of $0$ to $100$ meters above the sea level) 
equipped with clocks and measuring devices for gravitational field, 
and all the matter fields coupling to it, and their rates of change 
as to have initial data at a certain time slice $\Sigma$ for the system consisting of gravity and gravitating matter. 
From those measurements 
we should be able to predict (and retrodict) physical properties of the fields 
inside the domain of dependence of that hypersurface $D(\Sigma)$. 
In order to start calculations we choose a coordinate chart for a spacetime region containing $D(\Sigma)$ 
and also choose a gauge. 
Then the measurements are translated into functionals of the 
initial conditions for a system of partial differential equations that can be solved inside $D(\Sigma)$. 
In fact, one thing that can be calculated is the actual location of $D(\Sigma)$ according to the chosen coordinate chart. 
Let us call $U$ the spacetime domain resulting from 
a truncation of $D(\Sigma)$ with a topology of the type $\Sigma \times [0, 1]$ (where $\Sigma$ is a three dimensional ball). 
The domain $U$ 
is endowed with a 
foliation in which one of the leaves is a truncation of $\Sigma$.%
\footnote{
A truncation procedure starts with 
removing a tubular neighborhood $T$ of $\partial \Sigma$ from $\Sigma$, and consider 
$A = D(\Sigma \setminus T)$. Then consider the subset $B$ of $D(\Sigma)$ which is covered by leaves $\Sigma_t$ with 
$t \in [\epsilon , 1-\epsilon ]$. A truncation with the properties we seek is $U = A \cap B$. 
} 
Let us assume that the system, consisting of General Relativity coupled to matter, 
has a description in which active spacetime diffeomorphisms are a gauge symmetry. 
In this case, the location of $D(\Sigma)$ 
in the host four manifold is gauge dependent. 
Picturing $D(\Sigma)$ as a given portion of a four manifold implies either 
partially fixing the gauge or 
considering equivalence classes (somehow defined in terms of the fields themselves) 
which are the ones associated to $D(\Sigma)$ resulting in a formalism 
where active diffeomorphisms are no longer a gauge symmetry. 
The formalism described in Section \ref{ContextSect} and Section \ref{GaugeSect}, 
in particular the definition of gauge perturbations, 
includes these ideas at its core; from the outset we worked on a given spacetime domain $U$ 
(which may be a portion of spacetime and where gluing descriptions over such spacetime domains enjoys 
of interesting properties \cite{OCs}). 
Thus, one way of picturing the spacetime domains $U$ endowed with a foliation 
used in the main body of this article 
may be as truncated domains of dependence of a given hypersurface with boundary. 
However, this interpretation works only in a certain regime of fields whose induced causal structure 
is compatible with such an interpretation. 
Observables would have that interpretation only (and possibly be well defined only) for a given domain of 
fields. Working with locally defined observables is a crucial point of 
our framework. 

\section*{Apendix B}\label{AppendixB}

In this appendix the submanifold of the jet in which the field equation 
$E(L) = 0$ holds will be denoted by ${\mathcal E}_L \subset JY$; 
the space whose elements are solutions of the linearized field equation will be denoted by 
${\mathfrak F}_U$, those which are almost locally Hamiltonian by $\hat{\mathfrak F}_U^{\rm LH}$, and 
the locally Hamiltonian ones by ${\mathfrak F}_U^{\rm LH}$. 
Clearly ${\mathfrak F}_U^{\rm LH} \subset \hat{\mathfrak F}_U^{\rm LH} \subset {\mathfrak F}_U$.

Here we consider the case in which the curve of solutions $\phi_t$ 
with $\phi_0 = \phi$ and 
determined by a vector field $w$ in ${\rm Sols}_U$
which satisfies condition (\ref{X}) and does not satisfy 
condition (\ref{TrivialAtBdary}). 
Thus, $w$ satisfies the following two conditions: \\
(i) $(\tilde{\Omega}_L(w, z)[\phi] = d \tilde{\rho}^w (z))[\phi]$ \\
for every solution $\phi$ and every vector field $z$ in ${\rm Sols}_U$, and where $\tilde{\rho}^w$ is an operator which 
depends linearly on a vector field, 
acts as a differential operator in the field and it is 
valued in $n-2$ forms of $M$. \\
(ii) $w_\phi(p) \neq 0$ \\
for some solution $\phi$ and some $p \in \partial U$. 


Now we choose an 
almost locally Hamiltonian vector field $v$ in ${\rm Sols}_U$, 
which means that 
$\mathscr{L}_{v} \tilde{\Omega}_L( y, z) [\phi] = 
d\tilde{\sigma}^v ( y, z) [\phi]$ 
for every solution $\phi$ and every pair of vector fields $y, z$ in ${\rm Sols}_U$, and where $\tilde{\sigma}^v$ is an operator depending linearly on each of the vector fields and acting as a differential operator in the field. 
Condition (\ref{HOCs}) expressing the compatibility of $v$ with a current $\tilde{F}$ may be written as 
$(\mathscr{L}_{z}\tilde{F} + \tilde{\Omega}_L( v, z) ) [\phi] = 
d\tilde{\sigma}^F (z) [\phi]$ 
for every solution $\phi$ and every vector field $z$ in ${\rm Sols}_U$ and where the differential operator $\tilde{\sigma}^F$ satisfies 
$\tilde{\sigma}^v ( y, z) [\phi] = (- \delta 
\tilde{\sigma}^F + \tilde{\lambda}^F) ( y, z) [\phi]$ 
(with $d \tilde{\lambda}^F ( y, z) [\phi]= 0$) 
for every solution $\phi$ and every pair of vector fields $y, z$ in ${\rm Sols}_U$. 
Then 
\begin{equation}\label{LWF}
\mathscr{L}_{w} \tilde{F} [\phi] = - \tilde{\Omega}_L( v, w) [\phi] - 
d \tilde{\sigma}^F(w) [\phi] 
= d( - \tilde{\rho}^w(v) - \tilde{\sigma}^F(w) )[\phi] , 
\end{equation}
where we have used (i). 

The equation 
$\frac{d}{dt}|_{t=0} f_\Sigma[\phi_t] = \int_\Sigma \mathscr{L}_{w} \tilde{F} = 0$ is one scalar condition 
which can be studied with $v$ and $\tilde{\lambda}^F$ as unknowns. Clearly for generic values of 
$v$ and $\tilde{\lambda}^F$ the above equation does not hold. 
In this way we arrive to the desired conclusion that also in the case considered in this appendix observables constructed 
integrating observable currents suffice to separate points in the curve of solutions 
$\phi_t$ in a neighborhood of $\phi_0 = \phi$.

For a more detailed presentation see \cite{OCs}.

\section*{Conflict of interest}
There are no conflicts of interest associated with the publication of this work. 

\section*{Funding Statement}
This work was partially supported by grants PAPIIT-UNAM IN 109415, IN100218 and by a sabbatical grant by PASPA-UNAM.

\section*{Acknowledgments}
I would like to thank Igor Khavkine for important clarifications on the subject of local observables. 
I acknowledge correspondence and discussions about the subject of the article with 
Jasel Berra, Homero Díaz, Laurent Fridel, Alberto Molgado, Robert Oeckl, Michael Reisenberger, Charles Torre, Jos\'e A. Vallejo and 
Luca Vitagliano.

\bibliography{GaugeFromHolography}{}
\bibliographystyle{unsrt}

\end{document}